\documentstyle[11pt,osid,graphicx]{article}

\title 
{Quantum phase transition in spin systems studied through 
entanglement estimators} 
\author 
{A. Fubini\\{\footnotesize\it Dipartimento di Fisica 
dell'Universit\`a di Firenze, Via G. Sansone 1, I-50019 Sesto F.no (FI), 
Italy, MATIS-INFM and Dipartimento di Metodologie Fisiche e Chimiche 
(DMFCI) dell' Universit\`a di Catania, V.le A.~Doria 6, I-95125 Catania, 
Italy}\\
S. Haas and T. Roscilde$^*$ \\{\footnotesize\it Department of Physics and 
Astronomy, 
University of Southern California, Los  Angeles, CA 90089-0484, U.S.A.}\\
V. Tognetti  \\{\footnotesize\it Dipartimento di Fisica dell'Universit\`a di
Firenze, Via G. Sansone 1, I-50019 Sesto F.no (FI) and Istituto
Nazionale di Fisica Nucleare, Sezione di Firenze \&
tognetti@fi.infn.it} \\
P. Verrucchi \\{\footnotesize\it Istituto nazionale di Fisica della 
Materia INFM-CNR, sezione di Firenze,
Via G. Sansone 1, I-50019 Sesto F.no (FI)}\\
\vskip .2truecm
$*$ {\small Present address: {\it Max-Plank-Institut f\"ur Quantenoptik,
Hans-Kopfermann-Str.1, Garching, Germany.}}}

\begin{document}
\maketitle
\begin{abstract}
Entanglement represents a pure quantum effect involving  two or more
particles. Spin systems are good candidates for studying this
effect and its relation with other collective phenomena ruled by
quantum mechanics. While the presence of entangled states can be
easily verified, the quantitative estimate of this property is
still under investigation. One of the most useful tool in this framework
is the concurrence whose definition, albeit limited to $S=1/2$
systems, can be related to the correlators.
We consider quantum spin systems
defined along chains and square lattices, and described
by Heisenberg-like Hamiltonians: our goal is to clarify the 
relation between entanglement and quantum phase transitions, as well as 
that between the concurrence the and the specific quantum state of
the system.

\end{abstract}

\section{Introduction}

The occurrence of collective behavior in many-body quantum systems
is associated with classical and quantum correlations. The latter,
whose name is entanglement, cannot be accounted for in terms of
classical physics and represents the impossibility of giving a
{\it local} description of a many-body quantum state. Entanglement
is expected to play an essential role at quantum phase
transitions (QPT), where quantum effects manifest themselves at all
length scales, and the problem has recently attracted an increasing interest 
\cite{Osterlohetal02,Roscildeetal04,Roscildeetal05,OsborneN02,Vidaletal03,
Verstraeteetal04}.

Moreover, entanglement comes into play in quantum
computation and communication theory, being the main physical {\it
resource} needed for their specific tasks \cite{NielsenC00}. In
this respect, the perspective of manipulating entanglement by
tunable quantum many-body effects appears intriguing.

In this paper, entanglement estimators are found to give important
insight in the physics of $S=1/2$ spin systems, for which the {\it
concurrence} give quantitative definition of bipartite
entanglement. Quantum spin chains and two dimensional
lattices  in external fields are studied. Two striking features are
found: the occurrence of a factorized ground state at a field
$\,h_f\,$ and that of a QPT at $\,h_c>h_f\,$, where
multipartite entanglement plays an essential role.

The entanglement estimators have been calculated by
numerical simulations, carried out in models with linear
dimension $\,L\,$, via Stochastic Series Expansion (SSE) Quantum Monte
Carlo based on a modified directed-loop algorithm
\cite{SyljuasenS02}. We have verified that the inverse
temperature $\,\beta=2L\,$ is suitable to test the $\,T=0\,$
behaviour.

\section{The model}

We consider the antiferromagnetic $S=1/2$ XYZ model in a
uniform magnetic field:
 \begin{equation}\label{00}
  {\hat{\cal H}}/J =
   \sum_{\langle ij \rangle} \Big[ \hat{S}^x_i\hat{S}^x_{j}
   + \Delta_y \hat{S}^y_i\hat{S}^y_{j}
   + \Delta_z \hat{S}^z_i\hat{S}^z_{j}\Big] - \sum_i {\bf h}\cdot\hat{\bf S}_i~,
  \label{e.XYZhz}
 \end{equation}
where $J{>}0$ is the exchange coupling, $\langle ij \rangle$ runs
over the pairs of nearest neighbors, and ${\bf h}{\equiv}
g\mu_{\rm B} {\bf H}/J$ is the reduced magnetic field. The
canonical transformation $\hat{S}^{x,y}_i {\to}({-}1)^{I}
\hat{S}^{x,y}_i$ with $I=1(2)$ for $i$ belonging to sublattice
$1(2)$, transforms the coupling in the $xy$ plane from
antiferromagnetic to ferromagnetic. The Hamiltonian
(\ref{e.XYZhz}) is the most general one for an anisotropic
$S=1/2$ system with exchange spin-spin interactions. However, as
real compounds usually display axial symmetry, we will henceforth
consider either $\Delta_y{=}1$ or $\Delta_z{=}1$. Moreover, we
will apply the field along the $z$-axis, i.e. ${\bf h}{=}(0,0,h)$.

This paper focuses on the less investigated case
 $\Delta_z{=}1$, defining the XYX model in a field.
 Due to the non-commutativity of the Zeeman and the exchange
 term, for $\Delta_y {\neq} 1$ this model is expected
 to show a field-induced QPT  on any D-dimensional bipartite lattice,
 with the universality class of the D-dimensional Ising model
 in a transverse field~\cite{Chakrabartietal96}.
The two cases $\Delta_y{{<}}1$ and $\Delta_y{>}1$ correspond to an
easy-plane (EP) and easy-axis (EA) behavior, respectively. The
ordered phase in the EP(EA) case arises by spontaneous symmetry
breaking along the $x$($y$) direction, which corresponds to a
finite value of the order parameter $M^x$($M^y$) below the
critical field $h_{\rm c}$. At the transition, long-range
correlations are destroyed, and the system is left in a partially
 polarized state with field-induced magnetization
 reaching saturation only as $h{\to}\infty$. This picture has been verified so far in D=1 only, both
analytically~\cite{Dmitrievetal02}and
numerically~\cite{Roscildeetal04,Cauxetal03}.

Besides its quantum critical behavior, a striking feature of the
model Eq.~(\ref{e.XYZhz}) is the occurrence of an exactly
factorized ground state for a field $\,h_{\rm f}(\Delta_y)\,$
lower than the critical field $\,h_{\rm c}$. This feature was
previously predicted in Ref. \cite{KurmannTM82} for magnetic
chains, and its entanglement behaviour has been studied in
Ref.~\cite{Roscildeetal04}. Our QMC simulations have given 
evidence~\cite{Roscildeetal05} for a factorized ground state to occur also 
in D=2. We have then rigorously generalized the proof of factorization to
the most general Hamiltonian Eq.~(\ref{e.XYZhz}) on any 2D bipartite 
lattice. The proof will be soon reported elsewhere, but we here outline 
the essential findings in the following.
The occurrence of a factorized ground state is particularly surprising if 
one considers that we are
dealing with the $\,S=1/2\,$ case, characterized by the most
pronounced effects of quantum fluctuations. However, in the class
of models here considered, they are fully
uncorrelated\cite{KurmannTM82} at $h=h_{\rm f}$, thus leading to a
classical-like ground state. 
\section{Entanglement estimators}

In order to calculate the {\it entanglement of formation}
\cite{Bennettetal96} in the quantum spin system
described by Eq.~(\ref{e.XYZhz})
we make use of the {\it one-tangle} and of the
concurrence. The one-tangle \cite{Coffmanetal00,Amicoetal04}
quantifies the $T=0$ entanglement of a single spin with the rest
of the system. It is defined as:
\begin{equation}\label{aa}
\tau_1 = 4 \det\rho^{(1)}\,,\:\;\: \rho^{(1)} = (I + \sum_\alpha
M^{\alpha} \sigma^{\alpha})/2\,;
\end{equation}
$\rho^{(1)}\,$ is the one-site reduced density matrix, $M^{\alpha}
= \langle \hat{S}^{\alpha} \rangle$, $\sigma^{\alpha}$ are the
Pauli matrices, and $\alpha=x,y,z$. In terms of the spin
expectation values $M^{\alpha}$, one has:
\begin{equation}\label{bb}
\tau_1 = 1 - 4 \sum_\alpha (M^{\alpha})^2 .
\end{equation}
The concurrence \cite{Wootters98} quantifies instead the pairwise
entanglement between two spins at sites $i$ and $j$, both at zero and
finite temperature. For the model of interest, in absence of
spontaneous symmetry breaking ($M^x = 0$) the concurrence 
reads~\cite{Amicoetal04}
\begin{equation}
C_{ij}= 2~{\rm max}\{0,C_{ij}^{'},C_{ij}^{''}\}~,
\label{e.tauC}
\end{equation}
where
\begin{eqnarray}
C_{ij}^{'}
&=&g_{ij}^{zz}-\frac{1}{4}+|g_{ij}^{xx}-g_{ij}^{yy}|~,
\label{e.C1}\\
C_{ij}^{''}&=&|g_{ij}^{xx}+g_{ij}^{yy}|-
\sqrt{\left(\frac{1}{4}+g_{ij}^{zz}\right)^2-(M^z)^2}~,
\label{e.C2}
\end{eqnarray}
and $g_{ij}^{\alpha\alpha}=
\langle\hat{S}_i^\alpha\hat{S}_{j}^\alpha\rangle$ are the static
correlators.

One-tangle and  concurrence are related by the
Coffman-Kundu-Wootters (CKW) conjecture~\cite{Coffmanetal00},
recently proved by Osborne and Verstraete~\cite{OsborneV06},
stating that 
\begin{equation}\label{cc}
\tau_1{\geq}\tau_2{\equiv}\sum_{j{\neq}i}C_{ij}^2\,,
\end{equation}
which expresses the crucial fact that pairwise entanglement does
not exhaust the global entanglement of the system, as entanglement
can also be stored in $3$-spin correlations, $4$-spin
correlations, and so on. Therefore $n$-spin entanglement and
$m$-spin entanglement with $m{\neq} n$ are mutually exclusive and
this is really a unique feature of entanglement as a form of
correlation. The difference with classical correlations is evident.

Due to the CKW conjecture, the \emph{entanglement
ratio}~\cite{Roscildeetal04} $R{\equiv}\tau_2/\tau_1{<}1$
quantifies the relative weight of pairwise entanglement, and its
deviation from unity shows the relevance of $n$-spin entanglement
with $n{>}2$. Although indirect, the entanglement ratio is a powerful tool
to estimate multi-spin entanglement.

\section{The factorized state}

In the one dimensional system described by Eq.~(\ref{00}) the occurence
of an exactly factorized ground state for a field $h_{\rm f}(\Delta_y)$ 
lower than the critical field $h_{\rm c}$, was predicted in 
Ref.~\cite{KurmannTM82}.
In the case of the XYX model, this {\it factorizing field} is 
 \begin{equation}
 h_{\rm f} = \sqrt{2(1+\Delta_y)}\,.
 \end{equation}
 At $h = h_{\rm f}$ the ground state of the model takes a product
 form $|\Psi\rangle = \bigotimes_{i=1}^{N} |\psi_i\rangle $, where
 the single-spin states $|\psi_i\rangle$
are eigenstates of $({\bf n}_{1(2)} \cdot \hat{\bf S})$, where
${\bf n}_{1(2)}$ being the local spin orientation on sublattice 1 (2).
 Taking ${\bf n} = 
(\cos \phi \sin \theta,\sin \phi \sin \theta, \cos \theta)$,
 one obtains \cite{KurmannTM82}
 \begin{equation}
 \phi_1 = 0\,, \,\phi_2 = \pi\,,
\;\:\theta_1 = \theta_2 = \cos^{-1}\sqrt{(1+\Delta_y)/2}\, .
 \end{equation}
In particular, for $h = h_{\rm f}$ the spin orientation in the
quantum state is exactly the same as in the classical limit of the
model made of continuous spins with effective spin length 
$S_{\rm eff}=1/2$; this means that quantum fluctuations only set the 
length of the effective classical spin.
 The factorized state of the anisotropic model continuously connects
 with the fully polarized state of the isotropic model in a field
 for $\Delta_y =1$ and $h=2$.

In the two-dimensional case, we have found that for any value of
the anisotropies $\Delta_y$ and $\Delta_z$, there exists an
ellypsoid in field space
\begin{equation}
  \frac{h_{x}^2}{(1{+}\Delta_y)(1{+}\Delta_z)}
   {+}\frac{h_{y}^2}{(1{+}\Delta_y)(\Delta_y{+}\Delta_z)}{+}
   \frac{h_{z}^2}{(1{+}\Delta_z)(\Delta_y{+}\Delta_z)}=4
 \label{e.hfact}
\end{equation}
such that, when $\bf h$ lies on its surface, the ground state of
the corresponding model is factorized, $|\Psi\rangle {=}
\bigotimes_{i=1}^{N} |\psi_i\rangle $. The single-spin states
$|\psi_i\rangle$ are eigenstates of $({\bf n}_I \cdot \hat{\bf
S})$, ${\bf n}_I$ being the local spin orientation on sublattice
$I$. We will hereafter indicate with ${\bf h}_{\rm f}$ ({\it
factorizing} field) the field satisfying Eq.~(\ref{e.hfact}); at
${\bf h}{=}{\bf h}_{\rm f}$, the reduced energy per site is found
to be $\epsilon{=}{-}(1{+}\Delta_y{+}\Delta_z)/2$. In the
particular case of $\Delta_z {=} 1$ and ${\bf h}{=}(0,0,h)$, the
factorizing field takes the simple expression $h_{\rm f} {=}
2\sqrt{2(1{+}\Delta_y)}$.  As for the structure of the ground
state, taking ${\bf n}_I{=}(\cos \phi_I \sin\theta_I,\sin
\phi_I\sin\theta_I,\cos \theta_I)$, the analytical expressions for
$\phi_I$ and $\theta_I$ are available via the solution of a system
of linear equations.

The local spin orientation turns out to be different in the EP and
EA cases, being
\begin{equation}
\phi_1 {=} 0\,,\, \phi_2 {=}\pi\,, \:\:\theta_1 {=} \theta_2 {=}
\cos^{-1}\sqrt{(1{+}\Delta_y)/2}\,,
\end{equation}
 for $\Delta_y{<}1$, and
 \begin{equation}
\phi_1 {=}\pi/2\,,\, \phi_2 {=} {-}\pi/2\,,\:\:
 \theta_1 {=} \theta_2 {=}
\cos^{-1}\sqrt{2/(1{+}\Delta_y)}\,,
\end{equation}
 for $\Delta_y{>}1$.

 Despite its exceptional character, the occurrence of
 a factorized state is not marked by any particular anomaly
 in the experimentally measurable thermodynamic quantities. 
On the other hand, entanglement estimators are able to pin down 
the occurrence of such factorized states with high accuracy, as shown
in the following section.

\section{Entanglement and quantum phase transitions}

 The entanglement estimators display
 a marked anomaly at the factorizing field, where they
 clearly vanish, as expected.
 When the field is increased above $h_{\rm f}$, the
 ground-state entanglement
 has a very steep recovery, accompanied by the QPT at $h_{\rm c} > h_{\rm f}$.
 The system realizes therefore an interesting
 entanglement effect controlled by the
 magnetic field. This feature is associated with the many-body behavior of
 the system and it is shown in Fig.1 and Fig.2 for one
 and two-dimensional systems, respectively.

 \begin{figure}[t]
\centering
\includegraphics[width=100mm]{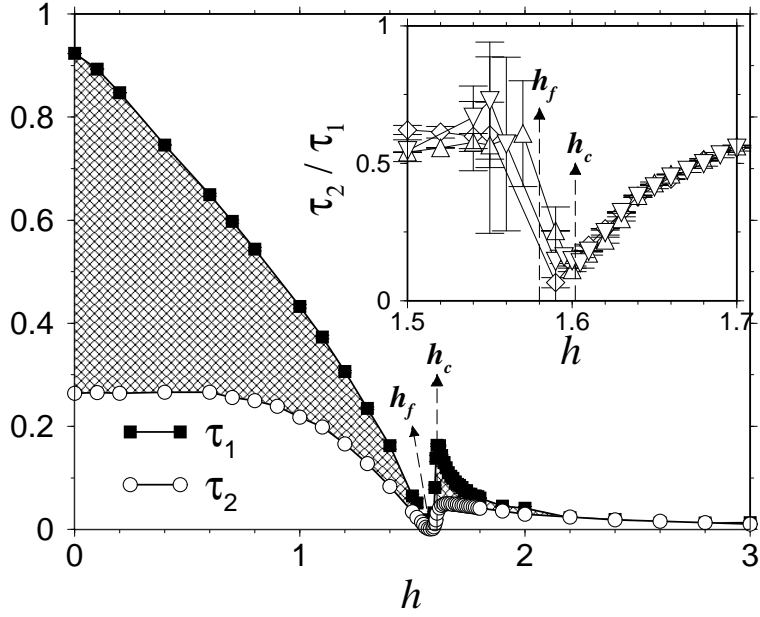}
\caption{One dimensional XYX model with $\Delta_y=0.25\,$: $\,\tau_1$
and $\tau_2$ versus $h$. Insets: entanglement ratio
$\tau_2/\tau_1$ close to the QPT. . \label{fig.1}}
\end{figure}

The concurrence terms, are generally short-ranged, and usually do not 
extend farther than the third neighbor.
The longest range of $C_{ij}$ is indeed observed around the
factorizing field $h_{\rm f}\,$\cite{Fubinietal06,Amicoetal06}.

The sum of squared concurrences $\tau_2$ is seen to be
 always smaller than or equal to the one-tangle $\tau_1$, both for one- 
and two-dimensional systems. This is in agreement with the CKW conjecture
 \cite{Coffmanetal00}. The total entanglement is
 only partially stored in two-spin correlations, and it appears also at the level of three-spin entanglement,
 four-spin entanglement, etc.
 In particular, we interpret the entanglement ratio
 \begin{equation}
 R = \tau_2/\tau_1  \,,
\end{equation}
as a measure of the fraction of the total entanglement stored in pairwise 
correlations. This ratio is plotted as a function of the field in the 
insets of both figures.
 A striking anomaly occurs at
 the quantum critical field $h_c$, where $R$ displays a very
 narrow dip. According to our interpretation, this result
 shows that the weight of pairwise entanglement decreases
 dramatically at the quantum critical point in favour of
 multi-spin entanglement. In fact, due to the CKW conjecture, and unlike 
classical-like correlations, entanglement shows the special property of 
{\it monogamy}, namely full entanglement
 between two partners implies the absence of entanglement with the
 rest of the system. Therefore multi-spin entanglement
 appears as the only possible quantum counterpart
 to long-range spin-spin correlations occurring at a QPT. This also 
explains the somewhat puzzling
 result that the concurrence remains short-ranged at a QPT while the 
spin-spin correlators become
 long-ranged \cite{Osterlohetal02}, and it evidences the
 serious limitations of concurrence
 as an estimate of entanglement at a quantum critical point.
 In turn, we propose the minimum of the entanglement
 ratio $R$ as a novel estimator of QPT, 
fully based on entanglement quantifiers.

\begin{figure}[t]
\centering
\includegraphics[width=100mm]{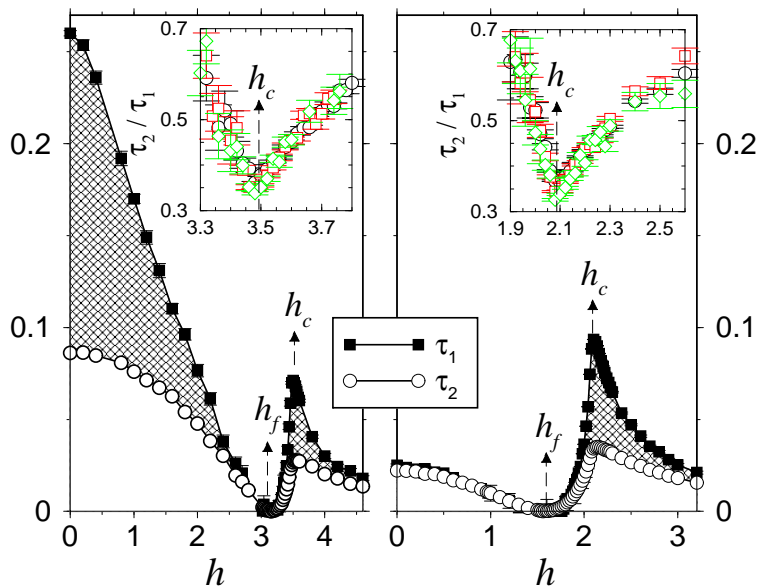}
\caption{Two dimensional XYX model: $\tau_1$ and $\tau_2$ versus
$h$. Left panel: easy-plane case ($\Delta_y =0.25$); Right panel:
easy-axis case ($\Delta_y =4$). Insets: entanglement ratio
$\tau_2/\tau_1$ close to the QPT.. \label{fig.2}}
\end{figure}

\section{From spin configurations to entanglement properties}

We here analyze the entanglement of formation between two spins,
as quantified by the concurrence $C$, in terms of spin configurations. 
In the simplest case of two isolated spins in the pure state 
$|\phi\rangle$ the concurrence may be written as $C=|\sum_i\alpha_i^2|$, 
where $\alpha_i$ are the coefficients entering the decomposition of 
$|\phi\rangle$ upon the magic basis. 
In this case, it can be easily shown~\cite{Fubinietal06} that $C$ extracts 
the 
information about the entanglement between two selected spins by combining 
probabilities and phases relative to some specific two-spin states.

In general, one should notice that a finite probability for two spins
to be in a maximally entangled (Bell) state does not guarantee \emph{per
se} the existence of entanglement between them, since this
probability may be finite even if the two spins are in a separable
state. In a system with decaying correlations, at infinite
separation all probabilities associated to Bell states attain the
value of $\,1/4\,$, but this of course tells nothing about the
entanglement between them, which is clearly vanishing. It is
therefore expected that differences between such probabilities,
rather than the probabilities themselves give insight in the
presence or absence of entanglement.

When the many-body case is tackled, the mixed-state concurrence of
the selected spin pair has an involved definition in terms of the
reduced two-spin density matrix \cite{Wootters98}.  We here assume
that ${\cal{H}}$ is real, has parity symmetry (meaning that either
${\cal{H}}$ leaves the $z$ component of the total magnetic moment
unchanged, or changes it in steps of $2$), and is further
characterized by translational and site-inversion invariance.

Let us select two specific spins in the system. We indicate by 
$p_\beta$, with $\beta=1,2,3,4$, the probabilities for the two spins
to be in one of the Bell states (maximally entangled), where $\beta=1,2$
(3,4) refers to the parallel (antiparallel) ones.
We do also introduce 
$p_{{_{\rm I}}}, p_{{_{\rm I\!I}}}, p_{{_{\rm I\!I\!I}}}, p_{{_{\rm
I\!V}}}$, which represent the probabilities for the two spins to be in
the (factorized) states of the standard base, where indexes $I$,$I\!I$
($I\!I\!I$,$I\!V$) refers to the parallel (antiparallel) ones.

It can be shown~\cite{Fubinietal06} that $\,C'\,$ and $\,C''\,$ 
(as defined in Eqs.~(\ref{e.C1}) and (\ref{e.C2}) with indexes dropped for
simplicity) can be written in terms of the above probabilities:
\begin{eqnarray}
2C'&=&|p_3-p_4|-2\sqrt{p_{{_{\rm I}}} p_{{_{\rm I\!I}}}}\label{e.C'p}~,\\
2C''&=&|p_1-p_2|-(1-p_1-p_2)=\nonumber\\
&=&|p_1-p_2|-2\sqrt{p_{{_{\rm I\!I\!I}}} p_{{_{\rm 
I\!V}}}}\label{e.C''p}~,
\end{eqnarray}
where we have used $p_{{_{\rm I\!I\!I}}}{=}p_{{_{\rm I\!V}}}$ and
hence $p_3+p_4=2p_{{_{\rm I\!I\!I}}}=2\sqrt{p_{{_{\rm
I\!I\!I}}}p_{{_{\rm I\!V}}}}$. The expression for $C''$ may be
written in the particularly simple form
\begin{equation}
2C''=2{\rm max}\{p_1,p_2\}-1~, \label{e.C''pmax}
\end{equation}
telling us that, in order for $C''$ to be positive, it must be
either $p_1>1/2$ or $p_2>1/2$. This means that one of the two
parallel Bell states needs to saturate at least half of the
probability, which implies that it is by far the state where the
spin pair is most likely to be found.

Despite the apparently similar structure of Eqs.~(\ref{e.C'p}) and
(\ref{e.C''p}), understanding $C'$ is more involved, due to the
fact that $\sqrt{p_{{_{\rm I}}}p_{{_{\rm I\!I}}}}$ cannot be
further simplified unless $p_{{_{\rm I}}}=p_{{_{\rm I\!I}}}$. The
marked difference between $C'$ and $C''$ reflects the different
mechanism through which parallel and antiparallel entanglement is
generated when time reversal symmetry is broken; ($p_{{_{\rm
I}}}\neq p_{{_{\rm I\!I}}}$ and hence $M_z\neq 0$). In fact, in
the zero magnetization case, it is $p_{{_{\rm I\!I}}}=p_{{_{\rm
I}}}=(p_1+p_2)/2$ and hence
\begin{equation}
2 C'=2{\rm max}\{p_3,p_4\}-1~, \label{e.C'p.Mz=0}
\end{equation}
which is fully analogous to Eq.~(\ref{e.C''pmax}), so that the
above analysis can be repeated by simply replacing $p_1$ and $p_2$
with $p_3$ and $p_4$.

For $M_z \neq 0$, the structure of Eq.~(\ref{e.C'p.Mz=0}) is
somehow kept by introducing the quantity
\begin{equation}
\Delta^2\equiv(\sqrt{p_{{_{\rm I}}}}-\sqrt{p_{{_{\rm I\!I}}}})^2~,
\label{e.Delta2}
\end{equation}
so that
\begin{equation}
2 C'=2{\rm max}\{p_3,p_4\}-(1-\Delta^2)~, \label{e.C'Delta}
\end{equation}
meaning that the presence of a magnetic field favors bipartite
entanglement associated to antiparallel Bell states, $|e_3\rangle$
and $|e_4\rangle$. In fact, when time reversal symmetry is broken
the concurrence can be finite even if $p_3,~p_4<1/2$.

>From Eqs.~(\ref{e.C''pmax}) and (\ref{e.C'Delta}) one can conclude
that, depending on $C$ being finite due to $C'$ or $C''$, the
entanglement of formation originates from finite probabilities for
the two selected spins to be parallel or antiparallel,
respectively. In this sense we speak about {\it parallel} and
{\it antiparallel} entanglement.

Moreover, from Eqs.~(\ref{e.C'p}) and (\ref{e.C''p}) we notice that,
in order for parallel (antiparallel) entanglement to be present in
the system, the probabilities for the two parallel (antiparallel)
Bell states must be not only finite but also different from each
other. Thus, the maximally entangled states result mutually exclusive in 
the formation of entanglement between two spins in the system, which is in 
fact present only if one of them is more probable than the others. The 
case $p_1=p_2=1/2$ ($p_3=p_4=1/2$)
corresponds in turn to an incoherent mixture of 
the two parallel (antiparallel) Bell states.

The above analysis suggests the first term in $C''$ ($C'$) to
distill, out of all possible parallel (antiparallel) spin
configurations, those which are specifically related with
entangled parallel (antiparallel) states. These characteristics
reinforce the meaning of parallel and antiparallel entanglement.

\section{Conclusions}

We have analyzed the behaviour of one- and two-dimensional $S=1/2$ 
antiferromagnets displaying a field-driven quantum phase transition. We 
have shown that while bipartite entanglement does not show any peculiar
feature at the transition, the entanglement ratio 
$R$~\cite{Roscildeetal04}, which measure the relevance of bipartite 
entanglement with respect to the total entanglement content of the system,
has a marked dip at criticality: this indicates that multipartite 
entanglement rules the QPT, $R$ being a powerful tool to detect it.

On the other hand, bipartite entanglement is found to efficiently detect
classical-like ground states, even the highly non-trivial ones which are
invisible under an analysis based upon standard magnetic observables.

Finally, an interpretation of the analytical expression of the concurrence
is given in terms of spin configurations, leading to a deeper insight into 
the relation between entanglement properties and state configurations in 
many-body systems.

\vfill
\eject

\end{document}